# Optimal Excitation Matching Strategy for Sub-Arrayed Phased Linear Arrays Generating Arbitrary Shaped Beams

P. Rocca, *Senior Member, IEEE*, L. Poli, *Member, IEEE*, A. Polo, *Member, IEEE*, and A. Massa, *Fellow, IEEE*

*Abstract*—The design of phased arrays able to generate arbitrary shaped beams through a sub-arrayed architecture is here addressed. The synthesis problem is cast in the excitation matching framework so as to yield clustered phased arrays providing optimal trade-offs between the complexity of the array architecture (i.e., the minimum number of control points at the sub-array level) and the matching of a reference pattern. A synthesis tool based on the k-means algorithm is proposed for jointly optimizing the sub-array configuration and the complex sub-array coefficients. Selected numerical results, including pencil beams with sidelobe notches and asymmetric lobes as well as shaped main lobes, are reported and discussed to highlight the peculiarities of the proposed approach also in comparison with some extensions to complex excitations of state-of-the-art sub-array design methods.

## I. INTRODUCTION

**P**HASED array antennas (*PA*s) are a key enabling technology for modern communications and radar systems [1][2]. Thanks to the high speed scanning, the easy reconfiguration, and the multi-function capability, they are suitable for a number of civil, commercial, and military applications of great interest including 5G [3][4], small-satellite communications [5], and anti-collision systems for autonomous driving [6]. Due to the increasing demand of lower costs towards the mass-market production, while fully-populated *PA*s (*FPA*s) turn out to be unaffordable architectures [7] since each element of the array is equipped with a transmit-receive module (*TRM*) including amplifiers and phase/time delays, unconventional phased arrays (*UPA*s) characterized by an irregular inter-element spacing and/or the aperiodic sub-arraying and/or the space-time modulation [8] have led to alternative solutions with good trade-offs between the complexity (i.e., the costs) and radiation performance.

Among *UPA*s, irregular sub-arrays have received a wider attention because of the capability of mitigating quantization and grating lobes, while still yielding a high aperture efficiency [9][10]. When designing an irregular sub-arrayed *PA*, two sets of degrees-of-freedom (*DoF*s) have to be determined: the membership of each array element to a cluster and the values of the beam-forming coefficients at the sub-array level. With reference to an $N$-element array with $Q$ sub-array ports, the cardinality of the solution space of all possible sub-array configurations amounts to $Q^N$. As a matter of fact, each element can be potentially grouped in one of the $Q$ clusters. It follows that testing all possible sub-array aggregations is computationally unfeasible also for arrays with few elements. For this reason, only small-scale clustered arrays have been originally dealt with by iteratively determining the sub-array coefficients through the pseudo-inversion of an over-determined system of linear equations for an *a-priori* given clustering [11]. Afterwards, nature-inspired optimization algorithms have been profitably adopted to exploit their effective sampling capabilities. For instance, the Simulated Annealing [12], the Genetic Algorithms [13][14], and the Differential Evolution [15][16] have been used. Hybrid strategies, which integrate global and local optimization techniques, able to synthesize optimal beam-forming weights of a clustered layout have been proposed [17][18], as well. However, despite the advanced exploration features supported by the growing computational resources and the capacity to prevent local minima (i.e., sub-optimal solutions) related to the non-convexity of the problem with respect to the sub-array configurations, the use of global optimizers is strongly penalized by the slow convergence rate. Therefore, once again, the design of clustered arrangements has been confined to small and medium size apertures.

An effective synthesis tool, named Contiguous Partition Method (*CPM*), for synthesizing the cluster layout and the

Manuscript received August 0, 2019; revised January 0, 2020.
This work has been partially supported by the Italian Ministry of Education, University, and Research within the Program "Smart cities and communities and Social Innovation" (CUP: E44G14000060008) for the Project "WATERTECH - Smart Community per lo Sviluppo e l'Applicazione di Tecnologie di Monitoraggio Innovative per le Reti di Distribuzione Idrica negli usi idropotabili ed agricoli" (Grant no. SCN_00489) and within the Program PRIN2017 (CUP: E64I19002530001) for the Project "CYBER-PHYSICAL ELECTROMAGNETIC VISION: Context-Aware Electromagnetic Sensing and Smart Reaction (EMvisioning)" (Grant no. 2017HZJXSZ), and the Project "Antenne al Plasma - Tecnologia abilitante per SATCOM (ASI.EPT.COM)" funded by the Italian Space Agency (ASI) under Grant 2018-3-HH.0 (CUP: F91I17000020005).
P. Rocca, L. Poli, A. Polo, and A. Massa are with the ELEDIA Research Center (ELEDIA@UniTN - University of Trento), Via Sommarive 9, 38123 Trento - Italy (e-mail: {paolo.rocca, lorenzo.poli, alessandro.polo.1, andrea.massa}@unitn.it).
P. Rocca is also with the ELEDIA Research Center (ELEDIA@XIDIAN - Xidian University), 3P.O. Box 191, No.2 South Tabai Road, 710071 Xi'an, Shaanxi Province - China (e-mail: paolo.rocca@xidian.edu.cn).
A. Massa is also with the ELEDIA Research Center (ELEDIA@L2S - UMR 8506), 3 rue Joliot Curie, 91192 Gif-sur-Yvette - France (e-mail: andrea.massa@l2s.centralesupelec.fr).
A. Massa is also with the ELEDIA Research Center (ELEDIA@UESTC - UESTC), School of Electronic Engineering, Chengdu 611731 - China (e-mail: andrea.massa@uestc.edu.cn).
A. Massa is also with the ELEDIA Research Center (ELEDIA@TSINGHUA - Tsinghua University), 30 Shuangqing Rd, 100084 Haidian, Beijing - China (e-mail: andrea.massa@tsinghua.edu.cn).

corresponding sub-array coefficients also for large arrays has been proposed in [19]. By casting the clustered array design within the excitation-matching framework and exploiting the Fisher's grouping theory [20], the cardinality of the synthesis problem has been reduced to the size of the space of the so-called contiguous partitions [19] equal to the binomial $\binom{N-1}{Q-1}$. Thanks to a suitable representation of the solution space, first through a non-complete binary tree and then as a direct acyclic graph [21], where each path codes a feasible/contiguous partition, the design of large sub-arrayed arrays has been made affordable. Such an approach has been further improved in terms of efficiency and success rate in reaching the optimal solution [22] with a customized integration of an Ant Colony Optimization. Furthermore, the design of sub-array architectures supporting multiple patterns has been addressed [23], as well. In all these works, the *CPM* has been used to synthesize the sub-array amplitudes of *PA*s along with the clustering configuration. More recently, the approach has been extended to optimize the sub-array phases [24][25]. However, either amplitude or phase excitations have been considered so far, but never both quantities together. Indeed, although the *CPM* guarantees the minimum - in the least-square sense - matching of the set of ideal/reference (i.e., independent for each element) excitations, these latter must be real-valued so that they can be ordered along a line [20].

This work presents a novel and optimal strategy for designing sub-array *PA*s that goes beyond such a limitation to allow the excitation-matching, still optimal in the least-square sense, of complex-valued reference weights and, thus, to enable the synthesis of sub-arrays affording arbitrary-shaped beam patterns [26]-[28]. Towards this end, first two extensions of the real-valued *CPM* to complex coefficients are presented to highlight the key concept of contiguous partition as well as the limitations of its more simple customizations and the need of a more proper definition of contiguity in the complex plane. Then, the *k-means* algorithm is exploited for clustering the array elements, while the sub-array coefficients are analytically derived [19][25]. The main motivation of using the *k-means* procedure is that it is a natural consequence of the Fisher's theory [20] for two-dimensional/complex domains as well as the high convergence rate [29].

The remaining of the paper is organized as follows. The mathematical formulation of the clustered synthesis is reported in Sect. 2, where the proposed design technique is also described still within the excitation-matching framework. A representative numerical analysis is then (Sect. 3) carried out through the presentation and the discussion of a set of selected synthesis results concerned with sub-arrayed arrays generating pencil beams with sidelobe notches and asymmetric lobes as well as shaped main lobes. Eventually, some conclusions and final remarks are drawn (Sect. 4).

## II. MATHEMATICAL FORMULATION

Let us consider a linear array of $N$ elements equally-spaced by $d$ along the $x$-axis to be grouped into $Q$ ($Q < N$) sub-arrays, each containing $N_q$ ($q = 1, ..., Q$) elements, so that $\sum_{n=1}^{N} N_q = N$ (Fig. 1). For beam-forming purposes, each $q$-th ($q = 1, ..., Q$) sub-array is fed by a *TRM* composed by an amplifier and a phase shifter providing a complex excitation $I_q$ (Fig. 1). The mathematical expression of the array factor of the beam generated at sub-array level turns out to be

$$AF(\theta) = \sum_{q=1}^{Q} I_q \sum_{n=1}^{N} \delta_{c_n q} e^{jk(n-1)d \sin \theta} \quad (1)$$

where $I_q = \alpha_q e^{j\varphi_q}$, $\alpha_q$ and $\varphi_q$ being the amplitude and the phase coefficients of the $q$-th ($q = 1, ..., Q$) sub-array, respectively. Moreover, $k = \frac{2\pi}{\lambda}$ is the wavenumber, $\lambda$ being the wavelength, $\theta$ is the angle measured from broadside, and $\delta_{c_n q}$ is the Kronecker delta function equal to $\delta_{c_n q} = 1$ if $c_n = q$ and $\delta_{c_n q} = 0$, otherwise. The integer vector $\mathbf{c} = \{c_n \in \mathbb{N} | 1 \leq c_n \leq Q : n = 1, ..., N\}$ univocally describes the membership of the $n$-th ($n = 1, ..., N$) array element to the $q$-th ($q = 1, ..., Q$) cluster.

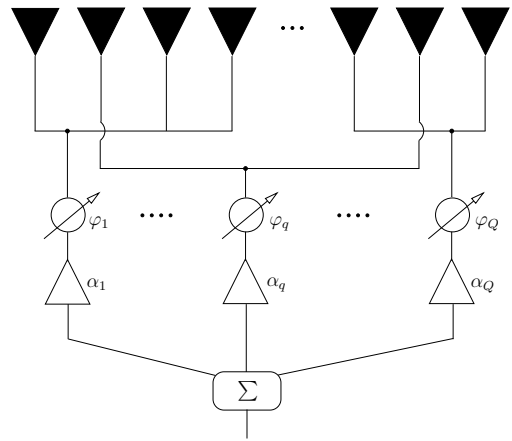

Figure 1. Sketch of the sub-arrayed architecture with sub-array level only complex-valued excitations.

According to this mathematical description, the synthesis problem at hand can be stated as follows:

***Sub-Arraying Synthesis Problem for Arbitrary Shaped Beams Generation*** - Given a set of complex excitation coefficients, $\{v_n; n = 1, ..., N\}$, generating a reference arbitrary-shaped pattern

$$AF^{ref}(\theta) = \sum_{n=1}^{N} v_n e^{jk(n-1)d \sin \theta}, \quad (2)$$

to determine the optimal clustering of the array elements into $Q$ disjoint sub-arrays, $\mathbf{c}^{opt} = \{c_n^{opt}; n = 1, ..., N\}$, and the values of the complex-valued sub-array excitations, $\mathbf{I}^{opt} = \{I_q^{opt}; q = 1, ..., Q\}$, so that the beam generated at sub-array level (1) is close as much as possible to the reference one (2).

Towards this aim, the synthesis problem is here reformulated as on optimization one in which the *DoF*s (i.e., the membership vector, $\mathbf{c}$, and the sub-array level beam-forming excitation vector, $\mathbf{I}$) are set to the values minimizing the following *pattern-matching* cost function

$$\Phi(\mathbf{c}, \mathbf{I}) = \frac{1}{\pi} \int_{\theta=-\pi/2}^{\pi/2} \left| AF^{ref}(\theta) - AF(\theta; \mathbf{c}, \mathbf{I}) \right|^2 d\theta. \quad (3)$$

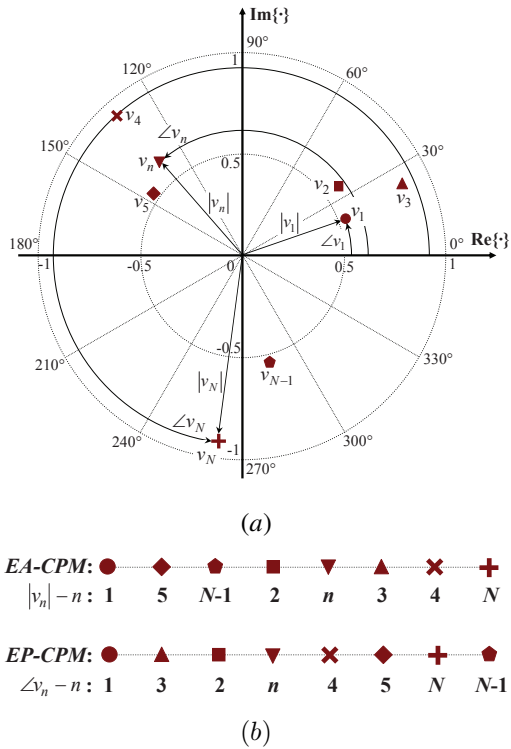

Figure 2. Reference excitations, $\{v_n; n = 1, ..., N\}$, (*a*) in the complex plain and (*b*) ordered on a list according to the *EA-CPM* and the *EP-CPM*.

By substituting (1) and (2) in (3), it follows that

$$\Phi(\mathbf{c}, \mathbf{I}) = \frac{1}{\pi} \int_{\theta=-\pi/2}^{\pi/2} \left| \sum_{n=1}^{N} \left( v_n - \sum_{q=1}^{Q} \delta_{c_n q} I_q \right) \times e^{jk(n-1)d \sin\theta} \right|^2 d\theta. \quad (4)$$

Thus, it turns out that the optimization of (3) is equivalent to the minimization of the following *excitation-matching* cost function

$$\Psi(\mathbf{c}, \mathbf{I}) = \frac{1}{N} \sum_{n=1}^{N} \left| v_n - \sum_{q=1}^{Q} \delta_{c_n q} I_q \right|^2 \quad (5)$$

since all the remaining terms in (4) are function of neither $\mathbf{c}$ nor $\mathbf{I}$. It is worth pointing out that (5) is a typical example of an over-determined problem with $N$ (complex) values to be approximated, in the least-square sense, by $Q$ ($Q < N$) ones. On the other hand, let us notice that the optimal value of the $q$-th ($q = 1, ..., Q$) sub-array coefficient, $I_q$, for a fixed clustering configuration $\mathbf{c}$ that minimizes (5) is the arithmetic mean of the reference excitations of the elements belonging to the same $q$-th ($q = 1, ..., Q$) cluster of the sub-arrayed architecture:

$$I_q(\mathbf{c}) = \frac{\sum_{n=1}^{N} \delta_{c_n q} v_n}{N_q}. \quad (6)$$

As a consequence, the synthesis can be performed by only looking for the best clustering configuration, $\mathbf{c}^{opt}$ ($\mathbf{c}^{opt} \triangleq arg\{min_\mathbf{c}[\Psi(\mathbf{c}, \mathbf{I})]\}$), since the sub-array weights come out as a free by product through (6) (i.e., $\Psi(\mathbf{c}, \mathbf{I}) \to \Psi(\mathbf{c})$).
As for the minimization of $\Psi(\mathbf{c})$, two methodological approaches are presented in the following. The former (Sect. *2.A*) is the most simple extension of the *CPM*-based approaches presented in [19] and [25], while the latter (Sect. *2.B*) is based on the generalization of the contiguity concept in clustering to the *2D* complex excitation plane.

### A. Complex-Extended CPM (E-CPM)

The extended *CPM* lies within the Fisher's grouping theory [20] and it requires the creation of an ordered list of (real) values. Once the list is defined, the Border Element Method (*BEM*), which has been proposed in [19] and further used in [25], is exploited for efficiently sampling the solution space of the possible clustering configurations (i.e., the contiguous partitions of the list of reference excitations). More specifically, the *E-CPM* works as follows:

- Step 0 - **Ordered-list Creation** - Given the set of $N$ complex values of the reference vector $\mathbf{v} = \{v_n; n = 1, ..., N\}$ [Fig. 2(*a*)], define the real-valued list $\mathbf{L} = \{\ell_n; n = 1, ..., N\}$ by ordering the reference excitations according to their magnitude (i.e., $l_1 = \min_n\{|v_n|\}$ and $l_N = \max_n\{|v_n|\}$) [19]. If two (e.g., the $m$-th and the $p$-th elements, being $m, p \in [1, N]$ and $m < p$) or more elements have the same magnitude, then order them according to their phase values (i.e., $l_m = \min\{\angle v_m; \angle v_p\}$ and $l_{m+1} = \max\{\angle v_m; \angle v_p\}$ subject to $|v_m| = |v_p|$) [Fig. 2(*b*)].
  Alternatively, unlike the previous amplitude-ordered version of the *E-CPM* (namely, the *EA-CPM*), the phase-ordered *E-CPM* (*EP-CPM*) defines the list $\mathbf{L}$ by ordering the coefficients according to their phase (i.e., $l_1 = \min_n\{\angle v_n\}$ and $l_N = \max_n\{\angle v_n\}$) [25], while they are ranked according to their magnitudes when the phase values are identical (i.e., $l_m = \min\{|v_m|; |v_p|\}$ and $l_{m+1} = \max\{|v_m|; |v_p|\}$ subject to $\angle v_m = \angle v_p$) [Fig. 2(*b*)];
- Step 1 - **Initialization** - Once $\mathbf{L}$ is defined, set the initial sub-array configuration, $\mathbf{c}^{(t)}$ ($t = 0$, $t$ being the iteration index), by randomly selecting $Q-1$ cut points among the $N-1$ admissible ones among the $N$ sorted coefficients $\{\ell_n; n = 1, ..., N\}$;
- Step 2 - ***BEM*-Based Solution-Space Sampling** - Execute the following steps:
  – Step 2.1 - **Sub-Array Coefficients Definition** - For the $t$-th trial sub-array configuration, $\mathbf{c}^{(t)}$, compute the optimal sub-array excitations $\mathbf{I}^{(t)}$ through (6);
  – Step 2.2 - **Cost Function Evaluation** - Compute the distance between the $Q$ sub-array coefficients and the $N$ reference ones by evaluating the excitation matching metric $\Psi^{(t)} = \Psi(\mathbf{c}^{(t)})$ (5);
  – Step 2.3 - **Solution Update** - Compare the current excitation-matching value $\Psi^{(t)}$ with the best cost function value found so far, $\Psi^{(t-1)}_{opt}$ ($\Psi^{(t-1)}_{opt} \triangleq \min_{h=0,...,t-1}\{\Psi^{(h)}\}$). If $\Psi^{(t)} < \Psi^{(t-1)}_{opt}$, then set $\Psi^{(t)}_{opt} = \Psi^{(t)}$ and update the current best clustering and the corresponding excitation vector: $\mathbf{c}^{(t)}_{opt} \leftarrow \mathbf{c}^{(t)}$ and $\mathbf{I}^{(t)}_{opt} \leftarrow \mathbf{I}^{(t)}$;
  – Step 2.4 - **Convergence Check** - Stop and go to Step 3 if the maximum number of iterations $T_{max}$ has

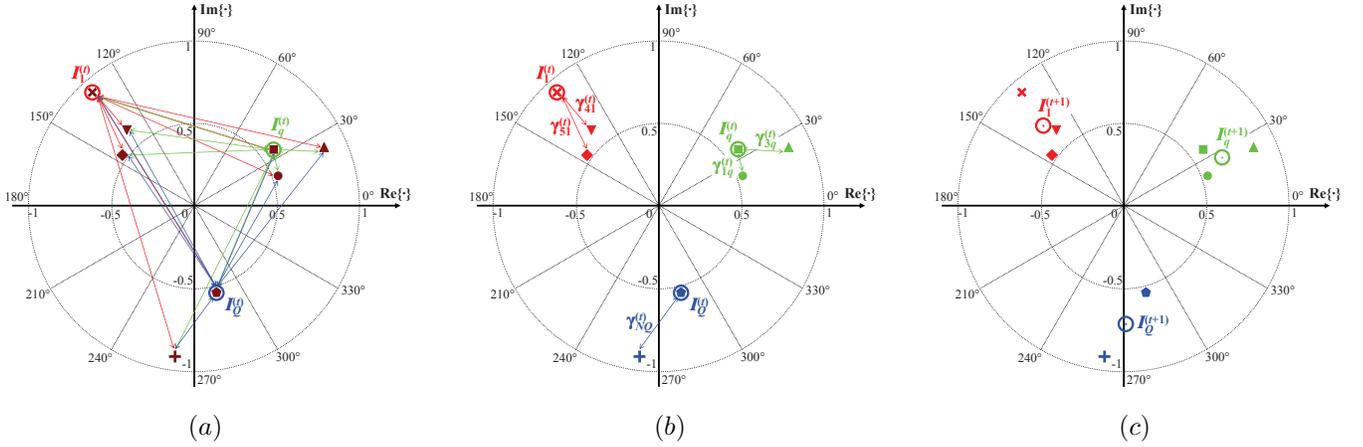

Figure 3. Diagrams related to (*a*) Step 2.1 - Distance Computation, (*b*) Step 2.2 - Element Clustering, and (*c*) Step 2.3 - Centroids Update of the *KMM*.

been reached ($t \geq T_{max}$) or the stationary condition

$$\frac{\left| T_{stat} \Psi_{opt}^{(t-1)} - \sum_{h=2}^{T_{stat}+1} \Psi_{opt}^{(t-h)} \right|}{\Psi_{opt}^{(t)}} \leq \eta \quad (7)$$

holds true, $T_{stat}$ and $\eta$ being user-defined parameters setting the number of iterations of the window for checking the stationary condition (7) and the minimum threshold for the decrease of the optimal value of the cost function, respectively. Otherwise, go to Step 2.5;

- Step 2.5 - **Sub-Array Configuration Update** - Update the iteration index ($t \leftarrow t+1$) and define the new sub-array configuration, $\mathbf{c}^{(t)}$, by changing the position of at least one of the $Q-1$ cut points of the previous sub-array partitioning, $\mathbf{c}^{(t-1)}$, according to the *BEM* procedure [19]. Then, go to Step 2.1;
- Step 3 - **Sub-Arrayed Array Design** - Set $\mathbf{c}^{opt} = \mathbf{c}_{opt}^{(t)}$ and $\mathbf{I}^{opt} = \mathbf{I}_{opt}^{(t)}$.

Although benefiting from the outcomes of the Fisher's grouping theory [20] and the effectiveness of the *BEM*, the *E-CPM* is a sub-optimal method since it casts the original complex-valued clustering problem into a real-valued one. In other words, ordering a set of complex values in terms of their amplitudes/phases is equivalent to project the corresponding representative points from a plane to a line, thus reducing the dimensionality (cardinality) of the solution space, , but also introducing an approximation. In order to fully deal with the *2D* nature of the clustering problem at hand, still keeping the principles of the Fisher's grouping theory, a second innovative method based on a customization of the *k-means* algorithm is described hereinafter (Sect. *2.B*).

### B. K-Means Method (KMM)

The minimization of the excitation matching metric (5), which is obtained with the association of $N$ array elements to $Q$ sub-arrays and the computation of the corresponding sub-array *centroids* through (6), can be mathematically classified as an unsupervised learning problem of divisive clustering [30] where a *2D* complex space of representative solution points has to be partitioned into $Q$ regions. Towards this end, the *k-means* algorithm is here customized according to the following procedural steps:

- Step 1 - **Centroids Initialization** - Initialize the centroids $\mathbf{I}^{(t)}$ ($t = 0$, $t$ being the iteration index) to $Q$ randomly-chosen complex-valued reference excitations among the $N$ available ones, $\{v_n; n = 1, ..., N\}$ [Fig. 3(*a*)];
- Step 2 - **KMM-Based Solution-Space Sampling** - Execute the following steps:
  - Step 2.1 - **Distance Computation** - For each $n$-th ($n = 1, ..., N$) reference coefficient, compute the Euclidean distance, $\gamma_{nq}^{(t)}$, from the $q$-th ($q = 1, ..., Q$) centroid of the set $\mathbf{I}^{(t)}$, $I_q^{(t)}$ [Fig. 3(*a*)]:

$$\gamma_{nq}^{(t)} = \left\| v_n - I_q^{(t)} \right\| = \\ \left[ \left( Re\{v_n\} - Re\left\{I_q^{(t)}\right\} \right)^2 - \\ \left( Im\{v_n\} - Im\left\{I_q^{(t)}\right\} \right)^2 \right]^{\frac{1}{2}} \quad (8)$$

  $Re\{\cdot\}$ and $Im\{\cdot\}$ being the real and the imaginary parts, respectively;
  - Step 2.2 - **Element Clustering** - Associate each $n$-th ($n = 1, ..., N$) array element to the $q$-th ($q = 1, ..., Q$) cluster (i.e., $c_n^{(t)} = q$) whose centroid has the minimum distance (8)

$$q = arg\left\{ min_{j \in [1, Q]} \left[ \left\| v_n - I_j^{(t)} \right\| \right] \right\} \quad (9)$$

  to create the $t$-th sub-array membership vector $\mathbf{c}^{(t)}$ [Fig. 3(*b*)];
  - Step 2.3 - **Centroids Update** - Update the iteration index ($t \leftarrow t + 1$) and compute the optimal (in the least-square sense) sub-array centroid vector $\mathbf{I}^{(t)}$ related to the clustering arrangement $\mathbf{c}^{(t-1)}$ with (6) [Fig. 3(*c*)];
  - Step 2.4 - **Convergence Check** - Stop and go to Step 3 if the maximum number of iterations $T_{max}$ has been reached ($t \geq T_{max}$) or $\mathbf{I}^{(t)} = \mathbf{I}^{(t-1)}$. Otherwise, go to Step 2.1;

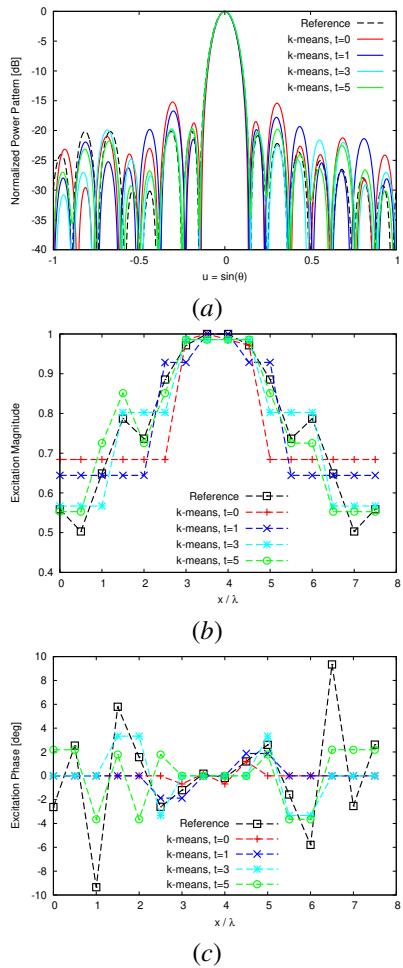

Figure 4. *Asymmetric Sidelobes Pencil Beam Pattern* ($N = 16$, $d = 0.5\lambda$, $Q = 4$) - Plot of the (*a*) power pattern, (*b*) the excitation amplitudes, and (*c*) the excitation phases of the *KMM* sub-arrayed solution at the iterations $t = \{0, 1, 3, 5\}$ along with the reference ones.

- Step 3 - **Sub-Arrayed Array Design** - Set $\mathbf{c}^{opt} = \mathbf{c}^{(t)}$ and $\mathbf{I}^{opt} = \mathbf{I}^{(t)}$.

It is worth pointing out that the execution of the steps from Step 2.1 up to Step 2.3 is equivalent to the minimization of the excitation matching metric, which turns out to be implicitly performed, without (5) being explicitly evaluated at each $t$-th iteration.

## III. NUMERICAL RESULTS

Representative results are presented and discussed in this Section to analyze the behavior of the proposed complex-valued clustering method and to assess its performance also in comparison with the *E-CPM*.

The first example deals with the design of a sub-arrayed array with $N = 16$, $d = \frac{\lambda}{2}$, and $Q = 4$ radiating a pattern close as much as possible to the reference one shown in Fig. 4(*a*) and characterized by asymmetric sidelobes. More in detail, the reference pattern has monotonically decreasing sidelobes on one side of the main beam ($0 < u \leq 1$, $u = \sin \theta$) from a level of $-20$ [dB] down to $-30$ [dB] along the end-fire direction [Fig. 4(*a*)]. On the other side ($-1 \leq u < 0$), a sidelobe depression of 10 [dB] with respect to a sidelobe level (*SLL*) of $-20$ [dB] is present within the angular range $-0.6 \leq u \leq -0.4$ [Fig. 4(*a*)]. The reference set of excitations, $\mathbf{v}$, has been computed by means of a Convex Programming optimization technique [31] and it is indicated with crosses $\times$ in the polar plots of Figs. 5(*a*)-5(*d*), while Figures 4(*b*)-4(*c*) show the corresponding amplitudes and phases, respectively.

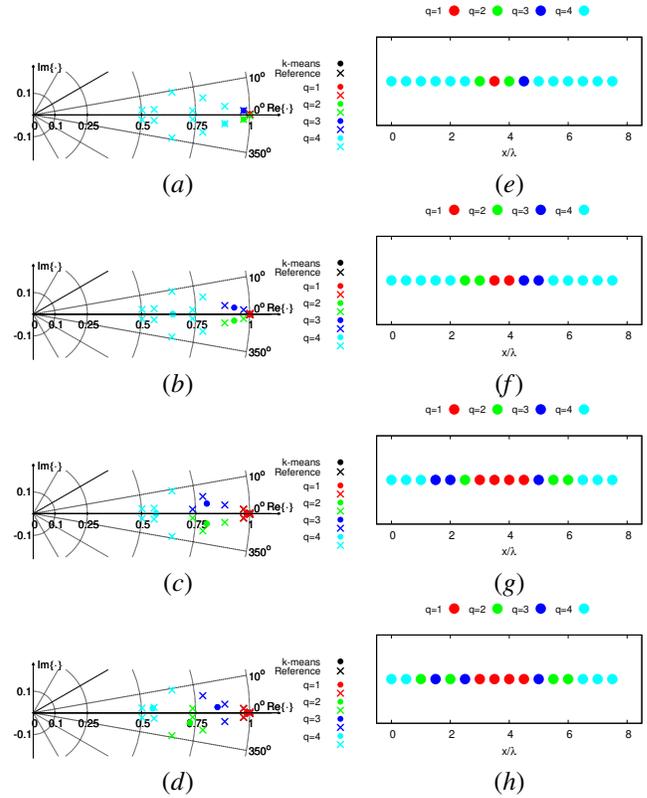

Figure 5. *Asymmetric Sidelobes Pencil Beam Pattern* ($N = 16$, $d = 0.5\lambda$, $Q = 4$) - Representation of (*a*)-(*d*) the reference, $\{v_n; n = 1, ..., N\}$, and the *KMM* sub-array, $\{I_q^{(t)}; q = 1, ..., Q\}$, excitations in the complex plane and (*e*)-(*h*) layout of the *KMM* clustered array at (*a*)(*e*) the initialization ($t = 0$), (*b*)(*f*) the first iteration ($t = 1$), (*c*)(*g*) the third iteration ($t = 3$), and (*d*)(*h*) the final iteration ($t = 5$).

Since the *KMM* is a local/deterministic searching method, its performance depends on the initialization, thus a set of $R = 50$ independent runs has been executed by considering different starting solutions. The evolution of the $Q$ sub-array excitations for a representative run converging to the median value of the excitation matching metric (5) among the $R$ executions is shown in Fig. 5. More specifically, the circles indicate the values of the $Q$ clustered excitations, while the crosses with the same color denote the reference weights used in (6) for the computation of the corresponding $q$-th ($q = 1, ..., Q$) centroid $I_q^{(t)}$ at the initial [$t = 0$ - Fig. 5(*a*)], the convergence [$t = 5$ - Fig. 5(*d*)], and two intermediate [$t = 1$ - Fig. 5(*b*); $t = 3$ - Fig. 5(*c*)] iterations. Moreover, the memberships of the array elements, which is coded into the vector $\mathbf{c}^{(t)}$, at the iterations $t = \{0, 1, 3, 5\}$ are shown in Figs. 5(*e*)-(*h*). According to the *KMM* implementation described in Sect. II.*B*, the centroids (i.e., the coefficients) of the $Q$ sub-arrays are randomly chosen at the initial iteration ($t = 0$) among the reference excitations.

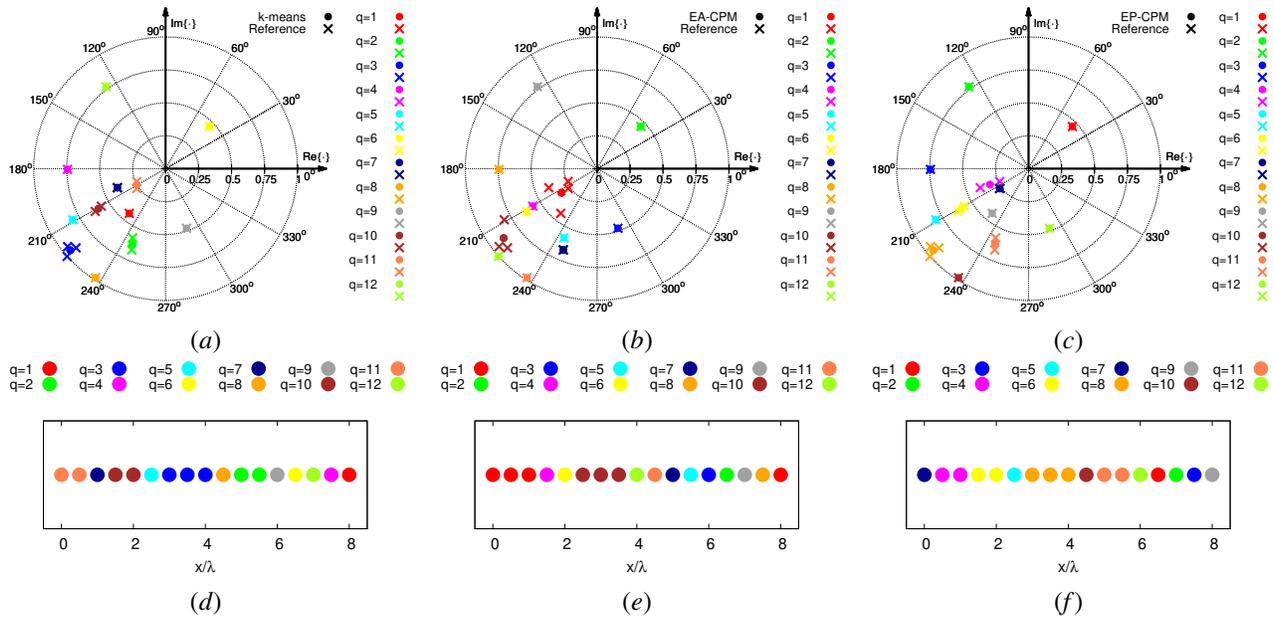

Figure 8. *Cosecant-Squared Beam Pattern* ($N=17$, $d=0.5\lambda$, $Q=12$) - Representation of (a)-(c) the reference, $\{v_n; n=1,...,N\}$, and the sub-array, $\{I_q^{(t)}; q=1,...,Q\}$, excitations in the complex plane and (d)-(f) layouts of the clustered arrays synthesized with the (a)(d) *KMM*, (b)(e) *EA-CPM*, and (c)(f) *EP-CPM*.

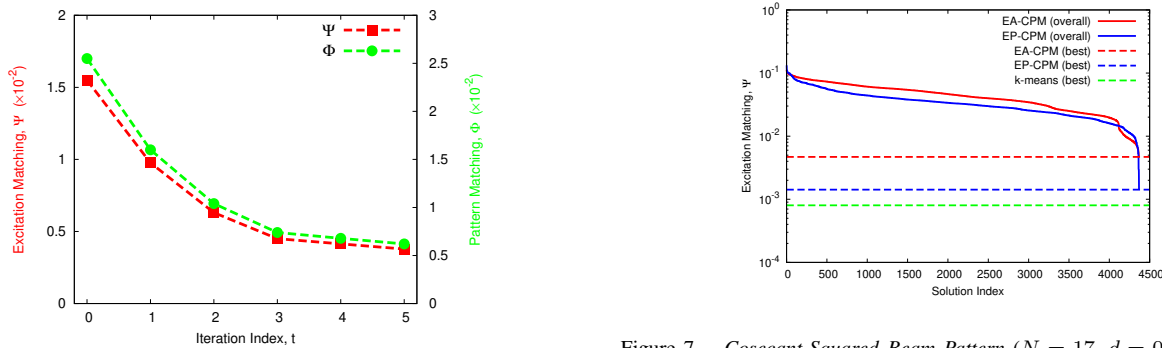

Figure 6. *Asymmetric Sidelobes Pencil Beam Pattern* (*KMM*; $N=16$, $d=0.5\lambda$, $Q=4$) - Behavior of the excitation matching, $\Psi^{(t)}$, and the pattern matching, $\Phi^{(t)}$, metrics versus the iteration index, $t$.

Figure 7. *Cosecant-Squared Beam Pattern* ($N=17$, $d=0.5\lambda$, $Q=12$) - Excitation matching values for the whole set of admissible solutions of the *EA-CPM* and the *EP-CPM* methods along with the optimal ones from the *KMM*, the *EA-CPM*, and the *EP-CPM*.

In this run, for instance, they have been set to $I_1^{(0)} = v_8$, $I_2^{(0)} = v_7$, $I_3^{(0)} = v_{10}$, and $I_4^{(0)} = v_6$ [Fig. 5(a)]. After associating each $n$-th ($n=1,...,N$) reference excitation to the closer $q$-th ($q=1,...,Q$) centroid (Step 2.2) [Fig. 5(a)], the sub-array configuration turns out to be as in Fig. 5(e), then the sub-array excitations are updated ($t=1$) through (6). The new positions of the centroids in the complex plane are shown in Fig. 5(b) and the corresponding clustering is given in Fig. 5(f). The clustering process stops after $t=5$ iterations when the stationary condition $\mathbf{I}^{(t)} = \mathbf{I}^{(t-1)}$ is reached. The optimal values of the sub-array excitations and the final clustering are shown in Fig. 5(d) and Fig. 5(h), respectively.

While the iterative procedure minimizes the excitation matching cost function (5), the beam generated at sub-array level better and better approximates the reference one as shown by the plots of the synthesized power patterns, $PP^{(t)}(\theta)$ ($PP(\theta) \triangleq |AF(\theta)|^2$) and of the reference one, $PP^{ref}(\theta)$, in Fig. 4(a). The same holds true for the corresponding sub-array amplitude [Fig. 4(b)] and phase [Fig. 4(c)] excitations. This behavior is also quantitatively highlighted in Fig. 6 since both the excitation matching $\Psi^{(t)}$ and the pattern matching $\Phi^{(t)}$ indexes monotonically decrease with the iteration thus confirming the effectiveness of the proposed method.

The second example is mainly devoted to perform a first comparison of the *KMM* with the *E-CPM* when synthesizing a shaped beam. Towards this end, the cosecant-squared pattern published in [32] has been selected as reference (see "*Reference*" in Fig. 9). It has been generated by a fully-populated array of $N=17$ elements equally-spaced by $d=\frac{\lambda}{2}$ and it exhibits a sidelobe level equal to $SLL^{ref} = -27.7$ [dB]. The clustered array to be designed is supposed to have $Q=12$ sub-arrays, that is $30\%$ saving of *TRM*s with respect to the fully-populated architecture. For comparison purposes, the best result from the *KMM*, reached $28\%$ times within $R=50$ independent runs, is considered along with the optimal solutions of the *EA-CPM* and the *EP-CPM*. Indeed, the size of the solution space of the *CPM* being equal to $\binom{N-1}{Q-1} = 4368$ contiguous partitions, all the admissible *CPM* solutions can be exhaustively generated, then the optimal one can be found by means of a simple enumerative procedure. The values of the excitation matching index (5) for every *CPM*-based solution, ordered from the worst (i.e., $\Psi\rfloor_{EA-CPM}^{worst} = 1.11 \times 10^{-1}$ and $\Psi\rfloor_{EP-CPM}^{worst} = 1.32 \times 10^{-1}$) to the best (i.e., $\Psi\rfloor_{EA-CPM}^{best} = 4.69 \times 10^{-3}$ and $\Psi\rfloor_{EP-CPM}^{best} = 1.42 \times 10^{-3}$) ones, are given in Fig. 7. On the same plot, the excitation matching value of the best synthesized *KMM* design (i.e., $\Psi\rfloor_{KMM}^{best} = 8.01 \times 10^{-4}$) is reported, as well. As it can be observed, the *KMM* outperforms all other methods since it reaches a solution with the minimum excitation matching value (Fig. 7). Moreover, it is worthwhile to highlight that such a *KMM* clustered arrangement [Figs. 8(*a*) and 8(*d*)] does not coincide with any of the contiguous partitions of the list of the sorted (amplitude/phase) reference excitations of the *EA-CPM* [Figs. 8(*b*) and 8(*e*)] or the *EP-CPM* [Figs. 8(*c*) and 8(*f*)].

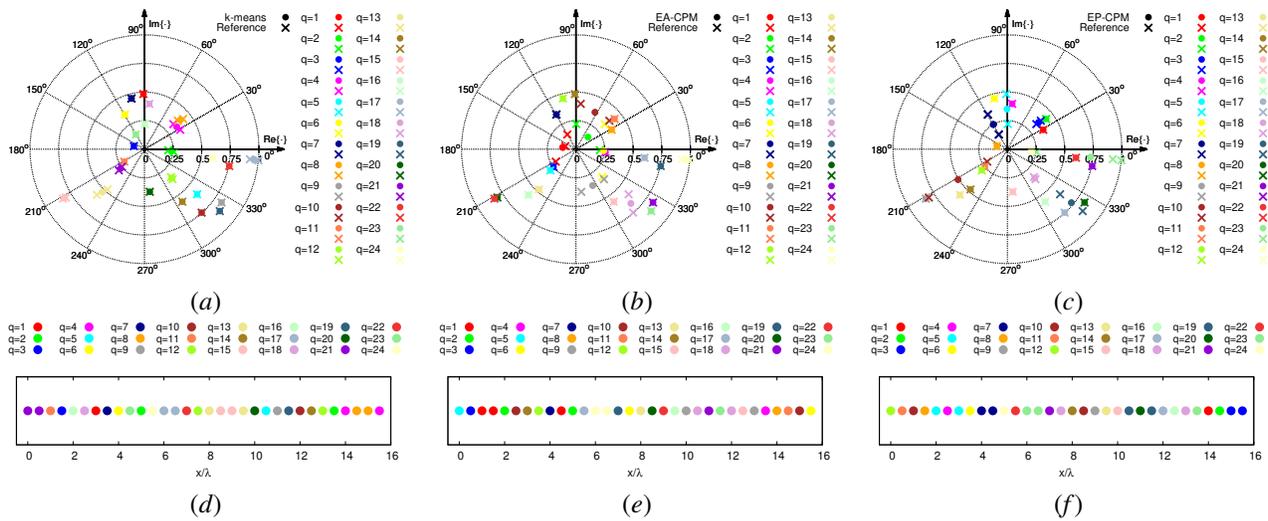

Figure 11.　*Flat-top Pattern* ($N = 32$, $d = 0.5\lambda$, $Q = 24$) - Representation of (*a*)-(*c*) the reference, $\{v_n; n = 1, ..., N\}$, and the sub-array, $\{I_q^{(t)}; q = 1, ..., Q\}$, excitations in the complex plane and (*d*)-(*f*) layouts of the clustered arrays synthesized with the (*a*)(*d*) *KMM*, (*b*)(*e*) *EA-CPM*, and (*c*)(*f*) *EP-CPM*.

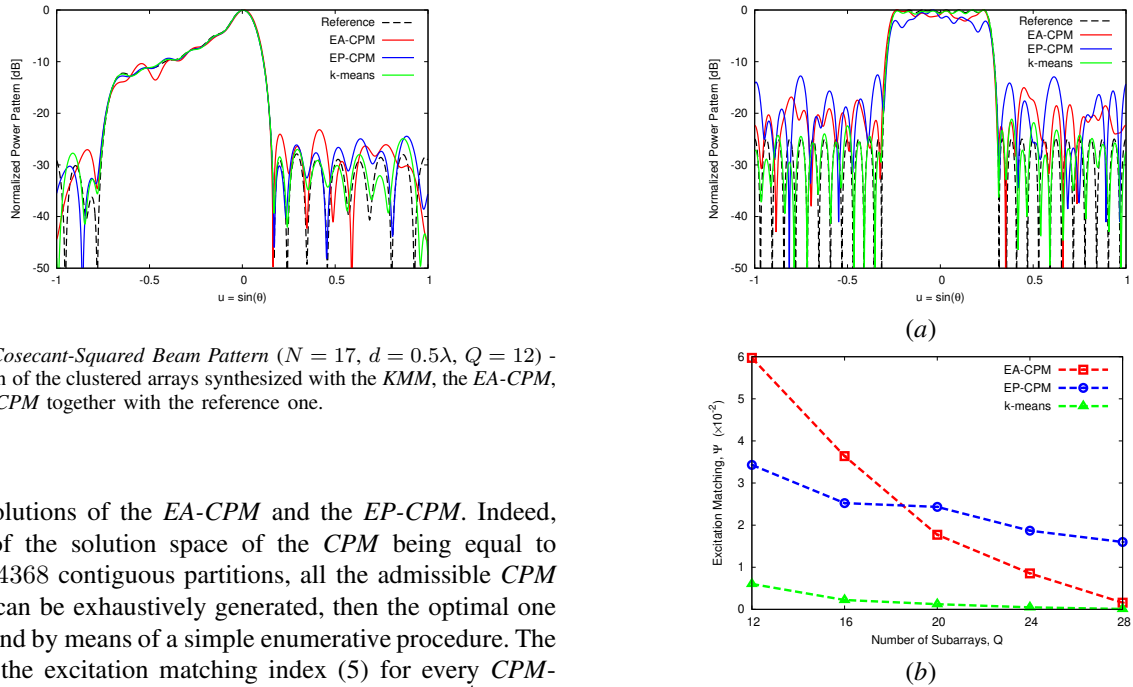

Figure 9.　*Cosecant-Squared Beam Pattern* ($N = 17$, $d = 0.5\lambda$, $Q = 12$) - Power pattern of the clustered arrays synthesized with the *KMM*, the *EA-CPM*, and the *EP-CPM* together with the reference one.

For completeness, Figure 9 shows the reference power pattern,

Figure 10.　*Flat-top Pattern* ($N = 32$, $d = 0.5\lambda$, $12 \leq Q \leq 28$) - Plot of the (*a*) power pattern of the sub-arrayed solutions synthesized with the *KMM*, the *EA-CPM*, and the *EP-CPM* methods when $Q = 24$ along with the reference one and (*b*) behavior of the excitation matching error, $\Psi^{opt}$, of the synthesized solutions versus the number of sub-arrays, $Q$.

$PP^{ref}(\theta)$, and those synthesized at sub-array level with the *KMM*, $PP^{KMM}(\theta)$, and the two *CPM*-based implementations [i.e., $PP^{EA-CPM}(\theta)$ and $PP^{EP-CPM}(\theta)$]. As expected from the values of the excitation matching index in Fig. 7, the *KMM* and the *EP-CPM* approximate the reference pattern better than the *EA-CPM*. For instance, the shape of the main lobe as well as the ripples of the secondary lobes are closer to the reference.

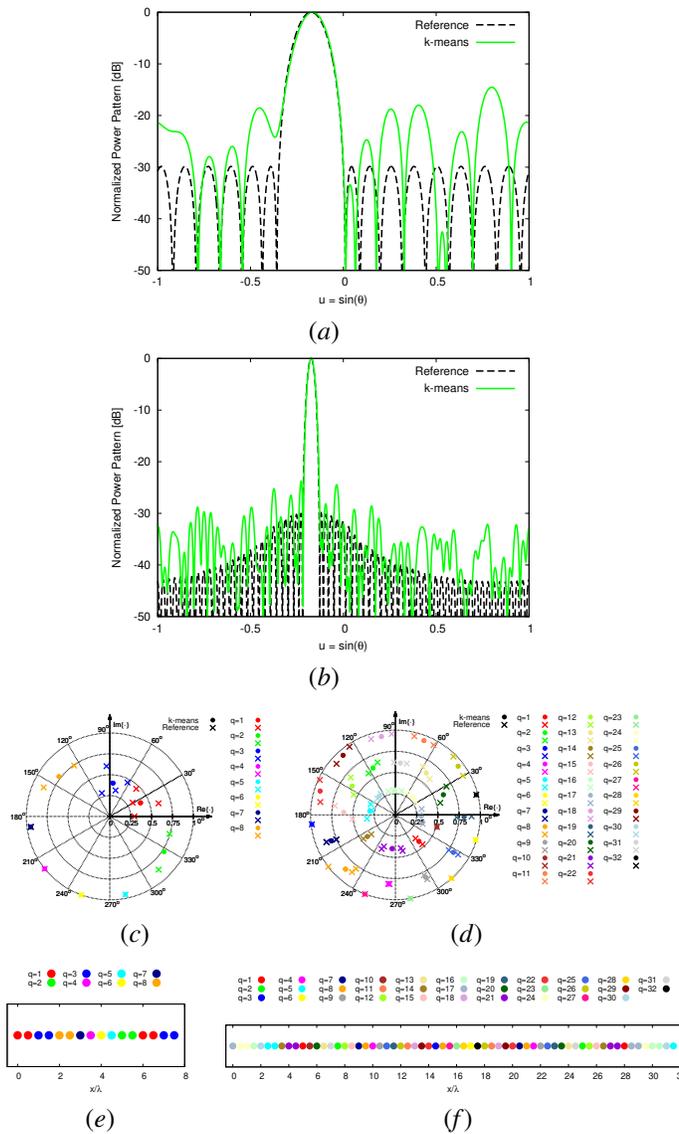

Figure 12. *Steered Pencil Beam Pattern* ($d = 0.5\lambda$, $Q = N/2$, $\theta_0 = -10$ [deg]) - Plot of the (a)(b) power pattern of the clustered solutions together with the reference ones, representation of (c)(d) the reference, $\{v_n; n = 1, ..., N\}$, and the sub-array, $\{I_q^{(t)}; q = 1, ..., Q\}$, excitations in the complex plane, and (e)(f) layouts of the clustered arrays synthesized with the *KMM* when (a)(c)(e) $N = 16$ and (b)(d)(f) $N = 64$.

The third example is concerned with a synthesis problem of higher complexity for which the exhaustive evaluation of all possible contiguous partitions is unfeasible. As a matter of fact, $\binom{N-1}{Q-1} = 84672315$ when $Q = 12$ and, thus, the *CPM* exploits the *BEM* for the solution space sampling. More in detail, a fully-populated array of $N = 32$ $\frac{\lambda}{2}$-spaced elements radiating a flat-top beam pattern with $SLL^{ref} = -25.0$ [dB] and having maximum main lobe ripples equal to $0.5$ [dB] has been taken into account ["*Reference*" in Fig. 10(a)]. By varying the number of sub-arrays in the range $12 \leq Q \leq 28$, the *KMM* always outperforms the *CPM*-based methods since it provides the best matching index whatever the number of clusters [Fig. 10(b)]. For illustrative purposes, the sub-array weights [Figs. 11(a)-11(c)] and the sub-array arrangements [Figs. 11(d)-11(f)] of the best solution yielded with the *KMM*

Table I
*Steered Pencil Beam Pattern* (*KMM* - $N = \{16, 32, 48, 64\}$, $d = 0.5\lambda$, $Q = N/2$, $\theta_0 = -10$ [DEG]) - VALUES OF THE EXCITATION MATCHING INDEX, $\Psi^{opt}$, THE PATTERN MATCHING INDEX, $\Phi^{opt}$, THE $SLL$, ALONG WITH THE COMPUTATIONAL COST, $\Delta\tau$.

|  | $\Phi^{opt}$ | $\Psi^{opt}$ | $SLL$ [dB] | $\Delta\tau$ [sec] |
|---|---|---|---|---|
| $N = 16$ | $5.50 \times 10^{-2}$ | $2.73 \times 10^{-2}$ | $-14.53$ | $0.109$ |
| $N = 32$ | $3.44 \times 10^{-2}$ | $1.69 \times 10^{-2}$ | $-19.41$ | $0.113$ |
| $N = 48$ | $2.08 \times 10^{-2}$ | $1.02 \times 10^{-2}$ | $-21.98$ | $0.117$ |
| $N = 64$ | $1.57 \times 10^{-2}$ | $7.71 \times 10^{-3}$ | $-23.77$ | $0.124$ |

[Fig. 11(a) and Fig. 11(d)], the *EA-CPM* [Fig. 11(b) and Fig. 11(e)], and the *EP-CPM* [Fig. 11(c) and Fig. 11(f)] are shown when $Q = 24$, while the reference power pattern together with the synthesized ones are reported in Fig. 10(a). Among the $R = 50$ runs, the success rate of the *KMM* in converging to the best solution varied in these cases between $2\%$ and $6\%$, with a drastic reduction with respect to the previous example due to the much higher cardinality of the solution space.

The efficiency of the *KMM* when designing larger arrays is assessed next by setting the reference beam to a Taylor pattern with $SLL^{ref} = -30$ [dB] and $\overline{n} = 7$ pointing along the direction $\theta_0 = -10$ [deg]. Four different apertures having $N = \{16, 32, 48, 64\}$ elements and inter-element spacing $d = \frac{\lambda}{2}$ have been considered, while keeping constant the ratio $\frac{Q}{N}$ to $\frac{1}{2}$. Table I summarizes the outcomes of this analysis by reporting the values of the excitation matching, the pattern matching, and the $SLL$ for the best *KMM* clustered solution running the code $R = 50$ times for each array size, $N$, with a success rate decreasing with the dimension of the solution space from $34\%$ for $N = 16$ to $2\%$ for $N = 64$. Although only half *TRM*s have been used as compared to the fully-populated layout, the matching with the reference pattern improves as $N$ is getting larger and larger since both $\Phi^{opt}$ and $\Psi^{opt}$ monotonically decrease and the arising $SLL$ better and better approximates the reference one. For illustrative purposes, the *KMM* solutions for $N = 16$ and $N = 64$ are shown in Fig. 12. Besides the power patterns [Figs. 12(a)-12(b)], the sub-array coefficients along with the reference excitations [Figs. 12(c)-12(d)] and the clustering of the array elements [Figs. 12(e)-12(f)] are reported for both array dimensions [$N = 16$ - Fig. 12(c) and Fig. 12(e); $N = 64$ - Fig. 12(d) and Fig. 12(f)]. As for the average *CPU*-time $\Delta\tau$ to synthesize a sub-arrayed arrangement, less than $0.13$ [sec] are required on a $2.4GHz$ PC with $2GB$ of RAM executing a non-optimized code whatever the aperture at hand (Tab. I).

## IV. CONCLUSIONS

The design of sub-arrayed *PA*s generating an arbitrary-shaped pattern has been addressed. Towards this aim, an innovative synthesis method, benefiting from the previously published excitation matching strategies and exploiting a clustering technique suitable for complex-valued excitations, has been proposed. The synthesis problem has been reformulated as an optimization one in which a customized version of the *k-means* has been used for defining the sub-array configuration of the array elements, while the complex-valued sub-array weights have been yielded in closed form as the arithmetic means of the reference excitations generating the target pattern and belonging to the same cluster.

The main methodological advances of this work with respect to the state-of-the-art can be summarized in the following ones:

- the theoretical formulation of the sub-array synthesis problem in the excitation-matching framework for effectively dealing with the sub-array level generation of arbitrary-shaped beams by extending the theory developed in [19] and [25];
- the introduction of an innovative and ad-hoc approach based on the *k-means* for solving the synthesis problem at hand.

From the numerical assessment, the proposed *KMM* design method proved:

- to overcome the limitations of the *CPM*-based methods by enabling the retrieval of sub-array configurations not achievable with the *EA-CPM* and the *EP-CPM*;
- to provide a high convergence rate and a significant computational efficiency regardless the number of array elements and sub-arrays;
- to enable the sub-array level synthesis of arbitrary-shaped beams, including pencil beams with asymmetric sidelobes as well as shaped main lobes (e.g., cosecant-square and flat-top beams).

However, it is important to observe that there is a trade-off between the capability of the proposed approach in matching the reference pattern and the complexity of the arising feeding network, characterized by sub-arrays also containing elements non-physically contiguous in the array aperture.

Future works, outside the scope and objectives of this paper, will be concerned with the development of a constrained version of the proposed approach guaranteeing the design of sub-arrays of physically contiguous elements, the integration of the proposed method with some power pattern synthesis technique taking advantage of the multiple solutions existing for the same power pattern shape in case of uniformly spaced antenna arrays, the design of clustered planar and conformal arrays as well as of non-uniformly spaced arrays, and the study of a global synthesis strategy instead of the use of a local one which, although very effective and robust, has performance still depending on the initialization.

ACKNOWLEDGEMENTS

A. Massa wishes to thank E. Vico for her never-ending inspiration, support, guidance, and help.